\documentstyle[12pt,epsf]{article}

\begin{document}   
   
\def\CA{{\cal A}}   
\def\CB{{\cal B}}   
\def\CC{{\cal C}}   
\def\CD{{\cal D}}   
\def\CE{{\cal E}}   
\def\CF{{\cal F}}   
\def\CG{{\cal G}}   
\def\CH{{\cal H}}   
\def\CI{{\cal I}}   
\def\CJ{{\cal J}}   
\def\CK{{\cal K}}   
\def\CL{{\cal L}}   
\def\CM{{\cal M}}   
\def\CN{{\cal N}}   
\def\CO{{\cal O}}   
\def\CP{{\cal P}}   
\def\CQ{{\cal Q}}   
\def\CR{{\cal R}}   
\def\CS{{\cal S}}   
\def\CT{{\cal T}}   
\def\CU{{\cal U}}   
\def\CV{{\cal V}}   
\def\CW{{\cal W}}   
\def\CX{{\cal X}}   
\def\CY{{\cal Y}}   
\def\CZ{{\cal Z}}   
   
\newcommand{\todo}[1]{{\em \small {#1}}\marginpar{$\Longleftarrow$}}   
\newcommand{\ads}[1]{{\rm AdS}_{#1}}   
\newcommand{\SL}[0]{{\rm SL}(2,\IR)}   
\newcommand{\cosm}[0]{\ell}   
\newcommand{\labell}[1]{\label{#1}\qquad_{#1}} 
\newcommand{\labels}[1]{\vskip-2ex$_{#1}$\label{#1}} 
\newcommand{\reef}[1]{(\ref{#1})}   
\newcommand{\tL}[0]{\bar{L}}   
\newcommand{\hdim}[0]{\bar{h}}   
\newcommand{\bw}[0]{\bar{w}}   
\newcommand{\bz}[0]{\bar{z}}   
\newcommand{\be}{\begin{equation}}   
\newcommand{\ee}{\end{equation}}   
\newcommand{\lp}{\lambda_+}   
\newcommand{\bx}{ {\bf x}}   
\newcommand{\bk}{{\bf k}}   
\newcommand{\tp}{\tilde{\phi}}   
\newcommand{\bound}[1]{{\partial {\cal M}}_{#1}}   
\newcommand{\volb}[1]{{\cal M}_{#1}}   
\newcommand{\lag}{{\cal L}}   
\newcommand{\fields}{\Phi}   
\newcommand{\B}{{\bf b}}   
\newcommand{\boundval}[1]{\bar{\fields}_{#1}}   
\newcommand{\K}{{\bf k}}  
\hyphenation{Min-kow-ski}   
   
\def\ie{{\it i.e.}}   
\def\eg{{\it e.g.}}   
\def\cf{{\it c.f.}}   
\def\etal{{\it et.al.}}   
\def\etc{{\it etc.}}   
   
\def\adv{{\it Adv. Phys.}}   
\def\ap{{\it Ann. Phys, NY}}   
\def\cqg{{\it Class. Quant. Grav.}}   
\def\cmp{{\it Comm. Math. Phys.}}   
\def\jetp{{\it Sov. Phys. JETP}}   
\def\jetpl{{\it JETP Lett.}}   
\def\jp{{\it J. Phys.}}   
\def\ijmp{{\it Int. J. Mod. Phys. }}   
\def\inc{{\it Nuovo Cimento}}   
\def\np{{\it Nucl. Phys.}}   
\def\mpl{{\it Mod. Phys. Lett.}}   
\def\pl{{\it Phys. Lett.}}   
\def\pr{{\it Phys. Rev.}}   
\def\prl{{\it Phys. Rev. Lett.}}   
\def\prcl{{\it Proc. Roy. Soc.} (London)}   
\def\rmp{{\it Rev. Mod. Phys.}}   
\def\dash{-----------------    }

\def\apr{\alpha'}   
\def\str{{str}}   
\def\lstr{\ell_\str}   
\def\gstr{g_\str}   
\def\Mstr{M_\str}   
\def\lpl{\ell_{pl}}   
\def\Mpl{M_{pl}}   
\def\varep{\varepsilon}   
\def\del{\nabla}   
\def\grad{\nabla}   
\def\tr{\hbox{tr}}   
\def\perp{\bot}   
\def\half{\frac{1}{2}}   
\def\p{\partial}   
   
\renewcommand{\thepage}{\arabic{page}}   
\setcounter{page}{1}

\rightline{HUTP-99/A016, EFI-99-9}   
\rightline{NSFITP-99-18, hep-th/9903190}   
\vskip 1cm   
\centerline{\Large \bf Spacetime and the}   
\vskip 0.4cm   
\centerline{\Large \bf Holographic Renormalization Group}   
\vskip 1cm   
   
\renewcommand{\thefootnote}{\fnsymbol{footnote}}   
\centerline{{\bf Vijay   
Balasubramanian${}^{1,2}$\footnote{vijayb@pauli.harvard.edu} and   
Per Kraus${}^{3}$\footnote{pkraus@theory.uchicago.edu},}}    
\vskip .5cm   
\centerline{${}^1$\it Lyman Laboratory of Physics, Harvard University,}   
\centerline{\it Cambridge, MA 02138, USA}   
\vskip .5cm   
\centerline{${}^2$ \it Institute for Theoretical Physics,}   
\centerline{\it University of California,}   
\centerline{\it Santa Barbara, CA 93106, USA}   
\vskip .5cm   
\centerline{${}^3$ \it Enrico Fermi Institute,}   
\centerline{\it University of Chicago,}   
\centerline{\it Chicago, IL 60637, USA}   
   
\setcounter{footnote}{0}   
\renewcommand{\thefootnote}{\arabic{footnote}}   
   
\begin{abstract}   
Anti-de Sitter (AdS) space can be foliated by a family of nested 
surfaces homeomorphic to the boundary of the space.  We propose a 
holographic correspondence between theories living on each surface in 
the foliation and quantum gravity in the enclosed volume. The flow of 
observables between our ``interior'' theories is described by a 
renormalization group equation.   The dependence of these flows on the 
foliation of space encodes bulk geometry. 
\end{abstract}   
   
\section{Introduction}   
  
The holographic principle~\cite{holog} states that quantum gravity on 
a manifold can be described by a theory defined on the boundary of 
that manifold.  The simplest realization of this principle has been in 
AdS space, which,  in certain cases, can be described by a {\em 
local} conformal field theory (CFT) defined on the AdS 
boundary~\cite{juanads}.  The correlation functions of the 
CFT describe the experiments of an observer who prepares 
field configurations at infinity and measures their amplitudes.

A strong version of the holographic principle would assert that 
quantum gravity on any volume contained within a manifold can be 
described by a theory defined on the boundary of that volume.  The 
holographic dual would then describe experiments of an observer who 
prepares field configurations on the interior boundary and measures 
their amplitudes. 
  
However, such an interior holographic dual within AdS cannot be a 
local theory.  To see this, assume that the  boundary 
theory is local, and that bulk objects near the boundary correspond 
to local excitations in the dual.  Then, imagine sending a light ray radially 
through $\ads{d+1}$, which has a metric: 
\begin{equation}  
ds^2= -(1+r^2/\ell^2) \, dt^2 + (1+r^2/\ell^2)^{-1} \, dr^2 +  
r^2 \, d\Omega^2_{d-1}.  
\end{equation}  
The time taken for a light ray to propagate from a point on the 
sphere at fixed $r$ to the antipodal point is: 
\begin{eqnarray}  
t_{{\rm bulk}} &=& 2 \int_0^r {dr \over 1 + r^2/\ell^2} = 2\ell   
\tan^{-1}{(r/\ell)} = \pi \ell - {2 \ell^2 \over r} + {\rm O}(r^{-2}) \\  
t_{{\rm bndy}} &=& {\pi r \over \sqrt{1 +r^2/ \ell^2}} = \pi \ell   
- {\pi \ell^3 \over 2 r^2} + {\rm O}(r^{-3}).  
\end{eqnarray}   
As noted in~\cite{light}, in the large $r$ limit 
the bulk and boundary propagation times are equal, indicating the 
potential consistency of a local holographic description.  But when 
$r$ is finite, $t_{{\rm bndy}} > t_{{\rm bulk}}$, so that nonlocal 
boundary dynamics will be necessary to yield the same arrival times. 
This also tells us that a holographic description of flat space should 
be nonlocal since taking $\ell \rightarrow \infty$ at fixed $r$ yields 
$t_{{\rm bndy}} = {\pi \over 2} t_{{\rm bulk}}$. 
  
In this note, we address the issue of interior holographic duals for 
AdS by adopting a Wilsonian renormalization group (RG) perspective.  To 
describe a subset of a system we ``integrate out'' the excluded 
degrees of freedom.  In general this will induce an infinite set of 
interactions in the remaining theory, making it nonlocal.  In the AdS 
context, we foliate spacetime by surfaces $\bound{\rho}$ of constant 
radial coordinate $\rho$, with enclosed volume $\volb{\rho}$.  We fix 
the values of the fields $\fields_\rho$ on $\bound{\rho}$ and perform 
the bulk path integral over the excluded volume.  The result is a 
nonlocal functional of $\fields_\rho$ which we treat as a boundary 
contribution to the bulk action describing $\volb{\rho}$.  Responses 
of the resulting interior path integral to variations of 
$\fields_\rho$ describe experiments carried out by observers placed on 
$\bound{\rho}$.  We identify these responses with the correlation 
functions of a holographic dual defined on the interior boundary. 
Related work has appeared recently in~\cite{PorSta}. For some other 
discussions of RG equations in the AdS/CFT context, see \cite{rgother}.

The observer at $\bound{\rho}$ naturally probes the interior volume 
with pointlike variations of the fields $\fields_{\rho}$.  In the 
semiclassical limit, the bulk equations of motion tell us that these 
variations turn into extended variations of the fields at infinity 
(see, {\it e.g.,} \cite{PeePol} and references therein).  This 
spreading of the fields  increases as $\bound{\rho}$ is 
moved into the interior.  In the CFT dual, these boundary values of 
bulk fields map onto sources smeared over a characteristic scale 
specified by the the position of the inner boundary.  It is then 
appropriate to integrate out  CFT degrees of freedom at lengths 
shorter than this scale.  This suggests that the interior holographic 
theories described above are related to the CFT duals of AdS spaces by 
coarsening transformations.  We will demonstrate that this is the case 
and show that, for any nested family of foliating surfaces for AdS, 
there is an RG equation describing the flow of observables in the 
corresponding series of interior holographic duals.  Spacetime 
diffeomorphisms relate foliating families and are realized as 
relations between different flows.

\section{Defining The Inner Correspondence}   
\label{sec:inner}

We will consider Euclidean AdS, which is topologically a 
ball.  Foliate AdS by a family of topologically spherical surfaces 
indexed by a parameter $\rho$ approaching $0$ at the boundary and 
$\infty$ at the center.  Let $\bound{\rho}$ be any element of this 
foliating family, with $\volb{\rho}$ being the enclosed volume.  The 
AdS/CFT correspondence for the boundary at $\rho=0$ is written 
as~\cite{gkp, holowit}: 
\begin{equation}   
e^{-Z_0[\Phi_0]} =   
\int_{\volb{0}} {\cal D}\fields \,    
e^{-S_0[\fields]}   
= \langle e^{-\int_{\partial\volb{0}}  \fields_0 \, {\cal O}} \rangle   
=   
e^{- S_{\rm CFT}(\fields_0)}   
\label{gkpw}   
\end{equation}   
The two terms on the left represent the string theory path integral on AdS  
 evaluated as a functional of   
the boundary data ${\fields}_0$.  On the right hand side is   
the effective   
action for the dual conformal field theory defined on the boundary   
manifold $\bound{0}$ in the presence of sources ${\fields}_0$.   
The spacetime action $S_0$ contains both bulk and boundary   
contributions:    
\begin{equation}   
S_0[\fields] = \int_{\volb{0}} \lag[\Phi] + \int_{\partial\volb{0}}  
B_0[\fields_0].   
\label{s0}   
\end{equation}   
where the boundary terms are chosen to cancel divergences arising from 
the bulk integral (see, e.g., \cite{counter}).  Upon 
performing the bulk path integral, $Z_0[\Phi_0]$ becomes a functional 
of $\Phi_0$ defined on $\bound{0}$.  Since the conformal factor on the 
boundary of AdS actually diverges, it is convenient to cut off the 
space at some small $\rho =\epsilon$, which can be understood as a 
kind of ultraviolet regulator for the CFT~\cite{susswitt}.  (We will 
always take $\epsilon \rightarrow 0$ in the end).  We write: 
\begin{equation}   
 Z_\epsilon[\Phi_\epsilon]  
= \sum_{n=1}^\infty    
\int_{\bound{\epsilon}}   
\left[ \prod_{j=1}^n \, d\B_j \, \sqrt{\gamma_\epsilon(\B_j)} \, \Phi_\epsilon(\B_j) \right] \, c_n(\epsilon;\B_1 \cdots \B_n)   
\label{kouter}   
\end{equation}   
Here $\B$ are boundary coordinates and $\gamma_\epsilon$ is the 
determinant of the induced metric on $\bound{\epsilon}$.  The boundary 
term $B_0$ in (\ref{s0}), when restricted to the surface 
$\bound{\epsilon}$, eliminates various contact terms that would 
otherwise make the expansion singular as $\epsilon \rightarrow 
0$~\cite{counter}.  The 
correlation functions of the dual CFT are precisely the coefficients 
$c_n$ in the $\epsilon \rightarrow 0$ limit.

We are interested in defining a suitable {\em inner correspondence} 
between quantum gravity on $\volb{\rho}$ and some theory defined on 
the boundary $\bound{\rho}$.  In the field theory limit we would like 
an equation analogous to (\ref{gkpw}): 
\begin{equation}   
e^{-Z_\rho[{\fields}_\rho]} =   
\int_{\volb{\rho}} {\cal D}\fields \,    
e^{-S_\rho[\fields]}   
=   
e^{- S_{\rm CFT}({\fields}_\rho)}  
.   
\label{anal}   
\end{equation}   
  
Consider an observer stationed on $\bound{\rho}$.  Such an observer can probe   
physics in the region $\volb{\rho}$ by measuring the amplitudes for various  
field configurations $\Phi_\rho$ to occur.  The amplitudes are given by the  
path integral in the full AdS spacetime subject to the boundary condition    
that $\Phi = \Phi_\rho$ on $\bound{\rho}$.  It is convenient to perform the  
path integral in two steps.  First, integrate over fields in the excluded  
volume $\volb{0} - \volb{\rho}$ to get a nonlocal functional of $\Phi_\rho$:  
\be  
e^{-Z_\rho[\Phi_\rho]} =   
 \int_{\volb{\rho}} {\cal D}\fields \,   
\int_{\volb{0} - \volb{\rho}} {\cal D}\fields \,    
e^{-S_0[\fields]}   
= \int_{\volb{\rho}} {\cal D}\fields \,e^{-S_\rho[\fields]}   
\ee  
where  
\be  
S_\rho[\fields]  = \int_{\volb{\rho}} \lag[\Phi] + \int_{\partial\volb{\rho}}  
B_\rho[\fields_\rho].  
\ee  
$S_\rho[\fields]$ encapsulates the physics in $\volb{\rho}$.     
  
The virtue of first integrating over the bulk fields in the excluded 
volume is that we can envision doing the analogous procedure in the 
gauge theory.  Roughly speaking, fields $\Phi_\rho$ correspond to 
smeared fields $\Phi_0$ at the outer boundary, and hence to smeared 
sources in the gauge theory.  In the CFT it is then natural 
to form an effective action by integrating over field modes with 
wavelengths shorter than the smearing length.  Matters will be made more 
concrete shortly when we consider the semiclassical limit. 
  
To compute bulk correlation functions on $\bound{\rho}$ we perform the  
remaining path integral over $\volb{\rho}$ to obtain a functional of  
$\Phi_\rho$,  
\begin{equation} 
Z_\rho[\Phi_\rho] = \sum_{n=1}^\infty \int_{\bound{\rho}}   
\left[   
\prod_{j=1}^n \, d\B_j \,   
\sqrt{\gamma_\rho(\B_j)} \,   
\Phi_{\rho}(\B_j)   
\right] \,   
c_n(\rho;\B_1 \cdots \B_n).  
\end{equation} 
We have obtained a one parameter set of correlation functions 
$c_n(\rho;\B_1 \cdots \B_n)$ indexed by $\rho$ which, by construction, 
reduce to those in (\ref{kouter}) as $\rho \rightarrow \epsilon$.  The 
dependence on $\rho$ is naturally interpreted as the renormalization 
group evolution of the correlation functions.

\subsection{Semiclassical correspondence}   
   
In the semiclassical, small curvature, limit the bulk path integral 
for the ``outer correspondence'' (\ref{gkpw}) is dominated by its 
saddlepoints.  So, in the corresponding limit of the dual CFT, 
(\ref{gkpw}) becomes 
\begin{equation}   
e^{-S_{{\rm cl}}({\fields}_0)} = e^{-S_{{\rm CFT}}({\fields}_0)}.   
\end{equation}   
The left hand side is now simply the AdS classical action (\ref{s0}) 
evaluated as a functional of boundary data.  We might have expected 
multiple saddlepoints to contribute, but, in Euclidean signature, and 
at least in the nearly free limit, demanding regularity for 
supergravity fields uniquely specifies classical solutions given the 
boundary data.  This even applies to the metric when the boundary data 
specifies a conformal structure sufficiently close to the round one 
(see discussion and references in \cite{holowit}).  If we admit black 
hole spacetimes the topology at infinity is no longer a sphere --- 
there is also a circle whose periodicity must be chosen to achieve 
regularity of the solution at the origin.  Again, the boundary 
conditions uniquely select the classical solution.  Multiple solutions 
can exist for fixed boundary conditions if we admit different bulk 
topologies.  For the present we will neglect the matter of summing 
over these solutions since they do not arise in the analysis of pure AdS.

To define the ``inner correspondence'' in the field theory limit we 
simply integrated over the fields in the excluded volume $\volb{0} - 
\volb{\rho}$.  In the semiclassical limit this amounts to evaluating 
the action for a classical solution in the excluded volume with fields 
taking values $\Phi_\rho$ at the inner boundary.  Again, there is a 
unique solution in the bulk with the prescribed boundary conditions. 
  
We can compute the full bulk action associated with classical solutions and  
express it in terms of either the fields $\Phi_\epsilon$ or $\Phi_\rho$:  
\begin{eqnarray}  
S_{cl} &=& \sum_{n=1}^\infty    
\int_{\bound{\epsilon}}   
\left[ \prod_{j=1}^n \, d\B_j \, \sqrt{\gamma_\epsilon(\B_j)} \, 
\Phi_\epsilon(\B_j) \right] \, c_n(\epsilon;\B_1 \cdots \B_n)  
\label{actions}  
\\  
&=&\sum_{n=1}^\infty \int_{\bound{\rho}}   
\left[   
\prod_{j=1}^n \, d\B_j \,   
\sqrt{\gamma_\rho(\B_j)} \,   
\Phi_{\rho}(\B_j)   
\right] \,   
c_n(\rho;\B_1 \cdots \B_n).  
\nonumber 
\end{eqnarray}  
To derive an RG equation we must relate the correlation   
functions at $\rho$ to those at $\epsilon$.  Such a relation is found  
by noting, as above, that the classical fields $\Phi_\epsilon$ are 
uniquely specified by $\Phi_\rho$.  We display this    relation in 
terms of a propagator:   
\begin{equation}   
\Phi_\epsilon(\B) = \int_{\bound{\rho}} \sqrt{\gamma_\rho(\B^\prime)}\, G_{\epsilon\rho}(\B,\B^\prime) \,   
\Phi_\rho(\B^\prime).  
\label{inout}   
\end{equation}

\subsection{The meaning of our construction}   
   
The meaning of our construction is most easily grasped by considering 
two point functions in the inner and outer theories.  In the 
semiclassical limit, the outer CFT two point function for widely 
separated operators is computed from a classical bulk geodesic between 
two boundary points.  Our procedure for computing inner two point 
functions amounts to extending the geodesic between two interior 
points until they reach the outer boundary, and adding in the action 
for the excluded part of the trajectory.  Since the geodesics spread 
on the way from the interior boundary to the exterior, interior 
correlators at one separation are given by exterior correlators at a 
larger separation.  More concretely, consider AdS in Poincar\'e 
coordinates: 
\begin{equation}  
ds^2 =  {\ell^2 \over \rho^2} (d\rho^2 + d\B^2)  
\end{equation}  
Consider a scalar field in AdS in a representation of the conformal 
group with weight $\Delta$.   Disturbances of this field on the AdS 
boundary ($\rho = 0$) propagate to the surface at fixed $\rho$ via the 
kernel   
\begin{equation}  
G_{{\rm bb}} \sim { \rho^{\Delta} \over (\rho^2 + |\B - 
\B^\prime|^2)^\Delta }    
\end{equation}  
So a point disturbance at $\bound{0}$ grows to a coordinate size $\rho$ at 
$\bound{\rho}$.  Conversely, a given point on $\bound{\rho}$ is 
affected by fields within a patch of coordinate size $\rho$ on the 
outer boundary.  Now imagine a observer on $\bound{\rho}$ who probes 
the system with local sources.   In terms of the original CFT, such an 
observer only has access to sources which are smeared over coordinate 
size $\rho$.   So her experiments  can be reproduced by an 
effective action in which degrees of freedom smaller than $\rho$ 
have been integrated out --- short distance information has been 
lost\footnote{If the observer can place sources on 
$\bound{\rho}$ with arbitrary precision, all CFT degrees of freedom 
must be retained.  This is because the effect of the smearing can be undone by 
making experiments at infinitesimal separations on $\bound{\rho}$. 
But of course such infinite precision experiments are beyond the 
validity of our supergravity analysis.}$\!$.  Integrating out degrees of 
freedom induces an infinite series of higher derivative terms,  
multiplied by powers of the dimensionful scale $\ell$.  If one tries 
to pass to the flat space limit by sending $\ell \rightarrow \infty$, 
the coefficients of the higher derivative terms diverge, signalling an 
increasingly nonlocal description. 
 
Our procedure realizes the argument of Susskind and 
Witten~\cite{susswitt} that the degrees of freedom in an interior 
holographic dual should scale like the boundary area.  Consider the 
surfaces $\bound{\rho}$ and $\bound{\rho^\prime}$ in Poincar\'e 
coordinates.  We have just argued that they are related to coarsenings 
of the CFT at infinity at scales $\rho$ and $\rho^\prime$.  Let us 
assume, following~\cite{susswitt}, that there is a fixed number of 
degrees of freedom per coarsened cell in the CFT at infinity.  Then the 
ratio of degrees of freedom in the theories on $\bound{\rho}$ and 
$\bound{\rho^\prime}$ is the ratio of areas, as desired.

An alternative procedure for defining the holographic dual of an 
interior volume is to simply cut off the interior path integral at 
some finite boundary.  This is unappealing because there are physical 
processes in which particles emerge from the interior region, 
propagate in the exterior, and then reenter the interior.  These 
processes are analogous to the virtual effect of massive degrees of 
freedom in a Wilsonian effective action.  In both cases, simply 
cutting off the theory throws out relevant physics.  In our approach, 
the effect of virtual processes is encoded in the nonlocal boundary 
terms.

\section{RG Flow of Observables}   
\label{sec:flow}   
   
We will now show that the flow of observables between our ``inner'' 
theories is described by a renormalization group equation.  As before, 
foliate Euclidean AdS by a family of surfaces homeomorphic to the 
boundary, and let $n^\mu$ be the outward pointing normal to this 
family of surfaces.  Then, if the spacetime metric is $g_{\mu\nu}$, 
the induced metric on a given foliating surface is $\gamma_{\mu\nu} = 
g_{\mu\nu} - n_\mu \, n_\nu$.  In an adapted coordinate system, with 
$\rho$ being the radial direction, the metric admits an ADM-like 
decomposition: 
\begin{eqnarray}   
g_{\mu\nu} &=& g_{\rho\rho} \, d\rho^2 + \gamma_{ij} \, (db^i + V^i \, 
d\rho) (db^j + V^j \,   d\rho) \\     
n^\mu &=& \delta^{\mu \rho} /\sqrt{g_{\rho\rho}}  \nonumber 
\end{eqnarray}   
Then, from  (\ref{actions}) and (\ref{inout}) we find that  
\begin{equation}   
c_n(\rho;\B_1\cdots\B_n)  =   
\int_{\bound{\epsilon}}    
\left[   
\prod_{j=1}^n d\B_j^\prime \sqrt{\gamma_\epsilon(\B_j^\prime)} \,    
G_{\epsilon\rho}(\B_j^\prime,\B_j)    
\right] c_n(\epsilon;\B_1^\prime\cdots\B_n^\prime)   
\label{qp}   
\end{equation}  
 We have just learned that the observables of the ``inner'' theory are 
precisely the ``outer'' CFT correlators convolved against the kernel 
$G_{\epsilon\rho}$.  To make progress, consider situations where we 
can undo the convolution by an integral transform.  For example, if 
the metric on $\bound{\epsilon}$ is proportional to the identity, the 
Fourier transform converts the convolution into a product. We will 
therefore refer to $c_n$ in the deconvolved basis as the ``momentum 
space'' correlator $\tilde{c}_n$, 
\begin{equation}   
c_n(\rho;\B_1\cdots\B_n) = \tilde{G}_{\epsilon\rho}(\B_1,\K_1) \cdots    
\tilde{G}_{\epsilon\rho}(\B_n,\K_n) \, \tilde{c}_n(\epsilon;\K_1\cdots \K_n)   
\end{equation}   
Here the variables $\K$ parametrize the deconvolution basis.  The 
correlator $\tilde{c}_n(\epsilon;\cdots)$ is independent of the index 
$\rho$ of the interior surface.  So the $\rho$ dependence of the inner 
observables is summarized by: 
\begin{equation}   
n^\mu  \grad_\mu \, c_n(\rho;\B_1\cdots\B_n) + \left[\sum_j n^\mu \grad_\mu \, \ln    
\tilde{G}_{\rho\epsilon}(\B_j,\K_j)\right] \, c_n(\rho;\B_1\cdots\B_n) = 0   
\label{eq:rggen}   
\end{equation}   
where $n^\mu$ is the normal vector to $\bound{\rho}$. 
(\ref{eq:rggen}) is an RG equation describing Wilsonian flow of 
correlators in the gauge theory, in correspondence with the 
observations of spacetime observers stationed on the fixed surfaces 
$\bound{\rho}$. The gradient operator acts on the coordinates $\B$ as 
well as on $\rho$.  This equation looks unfamiliar because it has been 
written for a general foliation of AdS.  We will see below that in 
Poincar\'e coordinates it displays all the expected features of 
Wilsonian renormalization group flow.

\subsection{Example: Poincar\'e coordinates}   
   
In Poincar\'e coordinates the metric of AdS is:   
\begin{equation}   
ds^2 = {\cosm^2 \over \rho^2} (d\rho^2 + d\B^2)   
\end{equation}   
and we are interested in surfaces of fixed $\rho$.  We will work out 
the relation between inner and outer observables for massive 
scalars. In $\ads{d+1}$ the operator dual to such a scalar has 
dimension $\Delta$: 
\begin{equation}   
\Delta = {d \over 2} + \nu, ~~~~~~~~~~~~~ \nu = {1 \over 2} \sqrt{d^2 
+ 4m^2}.   
\label{nudef}  
\end{equation}   
To Fourier transform both sides of (\ref{qp}) it is convenient to define the inner and outer correlators in momentum space:   
\begin{equation}   
\tilde{c}_n(\rho;\K_1\cdots\K_n) = \int_{\bound{\rho}} \left[ \prod_{j=1}^n d\B_j \,   
\sqrt{\gamma_\rho(\B_j)} e^{i\K_j\cdot\B_j} \right] \, c_n(\rho;\B_1 
\cdots \B_n)    
\end{equation} 
Next, since the propagator $G_{\epsilon\rho}$ approaches $\delta(\B - 
\B^\prime)/\sqrt{\gamma_\rho(\B)}$ as $\epsilon \rightarrow \rho$, the 
Fourier transform with respect to $\B^\prime$ gives 
$\tilde{G}_{\rho\rho}(\B,\K) = e^{i\K\cdot\B}$.  It is easy 
to construct a massive scalar mode solution that approaches such a plane 
wave on $\bound{\rho}$ from the complete bases provided in, {\it e.g.,}~\cite{bkl}. 
The propagator is then a Bessel function: 
\begin{equation}   
\tilde{G}_{\epsilon\rho}(\B,\K) = \left({\epsilon \over \rho}\right)^{d/2} \,   
\left( {K_\nu(q\epsilon) \over K_\nu(q\rho) } \right)   
\,   
e^{i\K\cdot\B}   
\end{equation}   
with $q^2 = \K \cdot \K$.  This gives:   
\begin{equation}   
\tilde{c}_n(\rho;\K_1\cdots\K_n) = \left( {\rho \over \epsilon} 
\right)^{-nd / 2} \,    
\left( \prod_j {K_\nu(q_j\epsilon) \over K_\nu(q_j\rho)} \right)   
\tilde{c}_n(\epsilon;\K_1\cdots\K_n)   
\label{eq:qppoin2}   
\end{equation}   
   
To gain further insight we need the power series expansion of the 
Bessel function:\footnote{The precise coefficients $e_\nu$ are not 
important for us.} 
\begin{eqnarray}   
K_\nu(z) &\propto& z^{-\nu}  [1 + F(z^2)] \\   
F(z^2) &=& \sum_{n=1}^\infty e_{-\nu}(n) \, z^{2n}   
- z^{2\nu} \, \sum_{n=0}^\infty e_{\nu}(n) \, z^{2n} \nonumber   
\end{eqnarray}   
(For purposes of argument we have have taken $\nu$ to be generic -- 
when $\nu$ is integral the expansion also involves logarithmic terms.) 
Using this in (\ref{eq:qppoin2}) we find: 
\begin{equation}   
\tilde{c}_n(\rho;\K_1\cdots\K_n) = \left({\rho \over \epsilon} \right)^{n (\Delta - d)} \,   
\left[ \prod_j {1 + F(q_j^2\epsilon^2) \over 1 + F(q_j^2\rho^2)} \right] \,   
\tilde{c}_n(\epsilon;\K_1\cdots\K_n)   
\end{equation}   
We implicitly understood all along that $\epsilon \rightarrow 0$. 
Since the theory is conformally invariant, this limit yields the 
scaling behaviour 
\begin{equation}   
\tilde{c}_n(\epsilon;\K_1\cdots\K_n) = \epsilon^{n(\Delta - d)} \, \bar{c}_n(\K_1\cdots\K_n)   
\end{equation}   
where $\bar{c}_n$ is finite.  Rearranging terms, the inner correlator 
becomes 
\begin{equation}   
\rho^{-n(\Delta - d)} \, \left[ \prod_j (1 + F(q_j^2\rho^2) \right] \, 
\tilde{c}_n(\rho;\K_1\cdots\K_n)     
=   
\bar{c}_n(\K_1\cdots\K_n)   
\label{eq:qppoin3}   
\end{equation}   
First consider $(q_j\rho) \ll 1$ for all $j$.  Then the interior 
correlators at $\rho$ and $\rho^\prime$ are related by a rescaling 
$(\rho/\rho^\prime)^{n(\Delta -d)}$.  This is exactly the behaviour 
expected for low energy correlation functions in a Wilsonian effective 
treatment.  We argued in Sec.~2 that the observables on the surface 
$\bound{\rho}$ in Poincar\'e coordinates were smeared at a scale 
$\rho$.  A Wilsonian treatment requires a rescaling of coordinates to 
keep the numerical size of the cutoff fixed.  Precisely this effect is 
achieved by the Weyl factor in the metric on $\bound{\rho}$ which 
keeps the proper size of the smearing fixed.  This in turn results in 
scaling of the correlators as we flow inwards (to the infrared). 
   
More generally, since $\bar{c}$ on the right hand side of 
(\ref{eq:qppoin3}) is independent of $\rho$, we have an RG equation: 
\begin{equation}   
\left[    \rho{\partial \over \partial \rho} - n(\Delta -d) 
\right]\tilde{c}_n(\rho;\K_1 \cdots \K_n)  
+  \left[\sum_j  \rho{\partial \over \partial \rho} \ln[1 + F(q_j^2 
\rho^2)]   
\right] \tilde{c}_n(\rho;\K_1 \cdots \K_n)   = 0 
\label{RGeqtn} 
\end{equation}   
When all the momenta $q$ are small, the second term vanishes and, 
as expected, we have the RG equation for pure scaling of infrared 
Wilsonian correlators. Violations of scaling appear in the second term
 and are suppressed at low momenta.  Transforming back into 
position space gives: 
\begin{equation}   
\left[\rho{\partial \over \partial \rho}  -n(\Delta -d)\right] \, c_n  + 
\left[\sum_j \rho   {\partial    
\over \partial \rho} \ln[1 + F(\rho^2\grad_i^2)]   
\right] \, c_n   = 0 
\end{equation}

\subsection{Bulk Field Equations from CFT?}

In the semiclassical limit, the interior effective theories that we 
have constructed are related to the exterior CFT by a renormalization 
group transformation, suggesting a direct relation between the bulk 
field equations and the RG equations in the CFT.  This is at first 
surprising since the bulk field equations are second order while the 
RG equations are first order.  However, there is no real conflict 
because demanding regularity of the bulk solutions in Euclidean space 
eliminates one solution, making the equations effectively first order. 
Related observations have been made in \cite{PorSta}. 
 
The connection can be made more explicit by recalling the 
correspondence between boundary behavior of the bulk fields in 
$AdS_{d+1}$ and sources and operators in the gauge 
theory~\cite{gkp,holowit,bkl,BDHM,bklt}.  Up to a $\rho$ dependent 
scaling, sources correspond to the boundary values of bulk fields 
while operators correspond to their radial derivatives. 
Schematically: 
\begin{equation} 
J \sim \Phi, \quad\quad\quad {\cal O} \sim \rho\, 
\partial_\rho \Phi. 
\label{corresp} 
\end{equation} 
In the CFT, $J$ appears as a coupling to the gauge invariant operator 
${\cal O}$ of the form $\int J({\bf b}) {\cal O}({\bf b})$.  In 
(\ref{corresp}), ${\cal O}$ is understood as the expectation value of 
the operator.  Now consider the structure of the bulk equation for a 
free scalar field of mass $m$: 
\begin{equation} 
(\Box - m^2) \, \Phi(\rho) \, e^{i\vec{k}\cdot \vec{x}} = 0 \quad 
\Rightarrow  \quad  
[\rho^2 \partial_\rho^2 + (1-d) \, \rho \partial_\rho - \vec{k}^2\, \rho^2 -m^2]  
\Phi(\rho) =0. 
\end{equation} 
If we use the relations (\ref{corresp}), the field equation takes the 
form 
\begin{equation} 
\left[\rho {\partial \over \partial \rho}  + d_0 \right]{\cal O}\, 
-\,\left[d_1+ d_2 \, \vec{k}^2 \, \rho^2\right] J =0. 
\label{fieldrg} 
\end{equation} 
Again, we are being schematic --- $d_0$, $d_1$ and $d_2$ are constants. 
The source $J$ is not an independent variable since it determines the 
expectation value for ${\cal O}$.  In momentum space, $J$ can be 
expressed as ${\cal O}$ times a function of $\vec{k}^2$.  Using this, 
we find that (\ref{fieldrg}) has the same form as  (\ref{RGeqtn}) with 
$n=1$. 
 
To make this connection precise, various issues such as the 
scheme dependence of the RG equations must be confronted. 
Nevertheless, there is reason to hope that the field equations of 
supergravity can be derived from the CFT via the renormalization group. 
Work in this direction is in progress.

\section{Discussion: Geometry and RG Flows}   
   
We have argued that there is a natural way to define an ``interior''  
holographic correspondence between physics inside finite volumes   
$\volb{\rho}$ and a theory on the boundary $\bound{\rho}$.  The   
correlation functions of the interior theory are related  to   
the exterior observables by a coarsening transformation.  A given   
family of foliating surfaces then leads to a particular flow of   
smeared observables summarized by a renormalization group equation.   
Changing the foliation leads to a different flow. In fact, we are   
learning that spacetime geometry arises in a holographic context as   
the geometry of the space of RG flows.

Consider a CFT defined on a plane and a family of theories derived 
from it by coarsening transformations.  Concretely, let $\phi(\B)$ 
be a field in the CFT, and  define coarsened fields $\phi(\rho;\B)$ by 
convolving $\phi$ against a kernel $K_\rho$ which has a characteristic 
scale $\rho$.  As $\rho$ increases from $0$ to $\infty$, we arrive at 
a family of smeared theories. In some natural sense there should be a 
geometry on this ``stack'' of theories.  First of all, a coarsening 
transformation should be accompanied by a rescaling of lengths, and 
that is implemented by rescaling the metric of the smeared theories. 
In addition, we would like a notion of distance or separation between 
the original CFT and its cousins that depends on the coarsening 
parameter $\rho$.  For the class of kernels inspired by AdS/CFT, we 
have learned that there is a natural distance, and it is given by the 
the geodesic length between the fixed $\rho$ Poincar\'e surfaces.  In 
this sense,  anti-de Sitter spaces induce a geometry on a certain 
class of RG flows of the dual CFTs.

In general there is no requirement that a field should be coarsened 
uniformly.  Indeed, it is often convenient in lattice field theory to 
use meshes of different sizes in different regions.  Within the 
AdS/CFT correspondence this freedom is realized in our ability to pick 
a general family of foliating surfaces for AdS space. 
Diffeomorphisms, or transformations between different families of 
foliating surfaces, are then manifestly realized as transformations 
between different RG flows.  The explicit action of diffeomorphisms on 
our  flows is easy to derive by acting with the 
generators on the kernel $G_{\epsilon\rho}$.

Several questions arise.  First, why is the class of smearing kernels   
selected by the AdS/CFT correspondence distinguished?  After all,   
there are many more ways to smear field variables than implied by   
solutions to AdS wave equations.  Second, is there a natural geometry   
on the space of RG flows that can be derived intrinsically from gauge   
theory considerations?  Given two theories with the same set of   
fields, we can detect the difference between them by computing and   
comparing correlation functions.  It is possible that there is a   
natural measure of distance between theories that can be derived in   
this way.  The answers to these questions are likely to be intimately   
related.  We are looking for a statement that some classes of   
coarsening transformations lead to RG flows on which there is a   
natural geometry.  That geometry, in a holographic context, is   
spacetime.

\vspace{0.2in} {\bf Acknowledgments:}    
 
We thank  Steve Giddings, David Gross, Gary Horowitz, Finn Larsen, Miao 
Li, Emil  Martinec, George Minic, Nikita Nekrasov, Joe Polchinski, and Ruud 
Siebelink for helpful discussions.    
V.B. is supported by the Harvard Society of Fellows and NSF grants   
NSF-PHY-9802709 and NSF-PHY-9407194.  P.K. is supported by    
NSF grant  PHY-9600697.


\end{document}